\documentclass[twocolumn,showpacs,preprintnumbers,amsmath,amssymb,prl]{revtex4-1}

\usepackage[T1]{fontenc}
\usepackage[utf8]{inputenc}
\usepackage{amstext}
\usepackage{amsmath}
\usepackage{graphicx}
\usepackage{esint}
\usepackage[english]{babel}
\usepackage{dcolumn}
\usepackage{bm}
\usepackage[FIGTOPCAP, hang, nooneline]{subfigure}
\usepackage[usenames,dvipsnames]{color} 
\usepackage{mathrsfs}
\usepackage{verbatim}
\usepackage{chemarr}
\usepackage[pdftex]{hyperref}
\usepackage{latexsym}
\usepackage[retainorgcmds]{IEEEtrantools}
 
\usepackage{pdfpages}

\newcommand{\tpsi}{\bar\psi}
\newcommand{\change}[1]{#1}
\newcommand{\rechange}[1]{#1}
\newcommand{\acomment}[1]{#1}
\newcommand{\enote}[1]{}
\newcommand{\anote}[1]{}
\newcommand{\ud}{\,\mathrm{d}}
\usepackage{float}

\begin{document}

\title{On~the~Validity~of~the~Law~of~Mass~Action~in~Three-Dimensional~Coagulation~Processes}

\author{Anton A. Winkler} \author{Erwin Frey}
\affiliation{Arnold Sommerfeld Center for Theoretical Physics and Center for NanoScience, Department~of~Physics, Ludwig-Maximilians-Universit\"at M\"unchen, Theresienstra{\ss}e 37, 80333 M\"unchen, Germany}

\begin{abstract}
Diffusion limited reactions are studied in detail on the classical coalescing process. We demonstrate how, with the aid of a recent renormalization group approach, fluctuations can be integrated systematically.  We thereby obtain an exact relation between the microscopic physics (lattice structure, particle shape and size) and the macroscopic decay rate in the law of mass action.  Moreover, we find a strong violation of the law of mass action. The corresponding term in the kinetic equations originates in long wavelength fluctuations and is a universal function of the macroscopic decay rate.
\end{abstract}

\acomment{\maketitle}

The law of mass action (LMA) is the fundamental law in chemical reaction kinetics. It states that the rate of an elementary reaction is proportional to the product of the concentrations of the participating molecules. 
In a seminal article that helped lay the foundations of a stochastic theory of chemical reaction kinetics, Smoluchowski provided a framework for the calculation of macroscopic decay rates and supported the validity of the LMA for three-dimensional systems~\cite{Smoluchowski:1917p16756,Chandrasekhar:1943p17335}. 
In the 1980s much effort was put in studying low dimensional systems, where it was found that strong correlations can lead to deviations from the LMA~\cite{Toussaint:1983p15598,Kang:1984p14320,Kang:1984p11987,Kopelman:1988p15594,Kuzovkov:1988p11649}.  
This anomalous behavior was observed, in particular, for the classical problem of coalescence, $A+A \rightarrow A$, where diffusing particles clot upon contact with a rate $\lambda$.  
By an approach designed for one dimension, one could even obtain exact solutions~\cite{BenAvraham:1998p15538}. This was complemented by results of the perturbative renormalization group at and below the critical dimension $d_c = 2$~\cite{Peliti:1986p11725,Lee:1994p10274}.
In contrast to this progress in low dimensions, advances for three-dimensional coagulation systems have remained largely elusive. In the experimental analysis, the LMA is still the `gold standard'~\cite{Zsigmondy:1917p16758,Avakian:1968p17336,Zhou:2010p16331}. \change{Indeed, it has obtained further support by field theoretic analysis proving the validity of Smoluchowski's heuristic arguments for asymptotically long times and low densities~\cite{Doi:1976p15932,Mikhailov}.} In this regime the density $\rho$ obeys the LMA rate equation $\partial_t \rho = - \mu \rho^2$, with a macroscopic decay rate $\mu$ which is a function of the microscopic rate $\lambda$ and of the size and shape of the particles. 

In this Letter we employ a non-perturbative renormalization group approach~\cite{Berges:2002p16426,Canet:2004p18293,Canet:2004p83,Canet:2006p16432} to study coagulation processes in three spacial dimensions. We find that the density obeys an equation of motion, $\partial_t \rho = - F(\rho)$, where the non-equilibrium `force' $F$ is derived from a non-equilibrium analog of a thermodynamic potential. It is obtained upon employing a non-perturbative renormalization group procedure, which successively integrates 
fluctuations starting at the microscopic scale. This provides a profound understanding of the intimate connection between the macroscopic description and the microscopic properties of the kinetic process. For low densities we recover the result from the law of mass action: $F(\rho) \approx \mu \rho^2$. An exact 
flow equation is derived which connects the microscopic rate $\lambda$ to the (non-universal) macroscopic rate $\mu$. Moreover, in contrast to previous work, the renormalization group approach enables us to explicitly incorporate both the effect of  lattice structure and of shape and size of the particles.
Lastly, we can calculate the non-equilibrium force $F(\rho)$ beyond the low-density limit. We find that long wavelength fluctuations give rise to a non-analytic term of the form  $c(\mu) \rho^{5/2}$. 
Remarkably, $c(\mu)$ is a simple, universal function of the macroscopic rate $\mu$. It depends on the microscopic features of the lattice and the particles only indirectly through its argument $\mu$.

\change{The coagulation process can be recast in terms of a field theory, 
an approach devised by several authors~\cite{Doi:1976p168,Zeldovich:1978p19619,Peliti:1985p17344}.} It has proven a powerful and versatile framework for reaction-diffusion processes in the past and also forms the basis of our non-perturbative renormalization group (RG) 
calculations. (For a simple example of the procedure, see Supplementary Material.) The stochastic dynamics corresponding to the reaction scheme $A+A\to A$ on a three-dimensional lattice is first mapped to its master equation and then translated to the action 
\begin{equation*}
	\rechange{S[\tpsi,\psi] = S_{\lambda}[\tpsi,\psi]  + S_\epsilon[\tpsi,\psi] + \int \! \mathrm{d}t \sum_\mathbf{x} \tpsi(\mathbf{x},t) \partial_t \psi(\mathbf{x},t) \, ,}
\end{equation*}
where the fields $\tpsi$ and $\psi$ are related to creation and annihilation of particles, respectively.  We wish to study particles that may extend over several lattice sites. This is achieved by introducing a reaction kernel $\lambda(\mathbf{y}-\mathbf{x})$ defining the rate at which a particle at site $\mathbf{x}$ annihilates another particle at site $\mathbf{y}$. The coagulation term $S_{\lambda} [\tpsi,\psi]$ then reads
\acomment{\begin{equation*} 
	 \int \! \mathrm{d}t  \sum_{\mathbf{x},\mathbf{y}} \lambda(\mathbf{y-x}) \left(\tpsi(\mathbf{y},t)+1\right)\tpsi(\mathbf{x},t) \psi(\mathbf{y},t)\psi(\mathbf{x},t) \,.
\end{equation*}}
While this determines the shape and size of the particles, the lattice structure is encoded in the diffusion term.
In Fourier space it is of the form
\acomment{\begin{equation*}
	S_\epsilon[\tpsi,\psi] = \int_{\mathbf{q},\omega} 
		 \epsilon(\mathbf{q}) \tpsi(-\mathbf{q},-\omega) \psi(\mathbf{q},\omega)  \, ,
\end{equation*}}
where $\int_\omega := \int \! \frac{\mathrm{d} \omega}{2 \pi}$, and $\int_{\mathbf{q}} := \int \! \frac{\mathrm{d}^3 q}{\left(2 \pi\right)^3}$ runs over the first Brillouin zone.  For concreteness, we consider a cubic lattice, where  the dispersion relation reads $\epsilon(\mathbf{q}) = 4 D \sum_{\nu = 1}^{3} \sin^2(q_\nu a/2)$~\cite{Dupuis:2008p12392}. We define time and length scale by setting the diffusion constant $D$ and the lattice spacing $a$ equal to 1.

The mean-field rate equation follows from the `classical field equations' given by the stationarity conditions $\delta S / \delta \psi = 0 = \delta S / \delta \tpsi$. 
The first equation is solved by setting the auxiliary field $\tpsi = 0$. 
Taking spatially homogeneous fields $\psi(\mathbf{x},t) \equiv \psi(t)$ and identifying $\psi$ with the density of particles ($\psi \to \rho$) yields the rate equation $\partial_t \rho =  - \sum_{\mathbf{x}}\lambda(\mathbf{x}) \rho^2$.  Hence, asymptotically the density behaves as $\rho(t) \sim {\left[\sum_{\mathbf{x}}\lambda(\mathbf{x})\right]}^{-1}  t^{-1}$.

To  account for  fluctuations one has to go beyond such a mean-field approach and consider the generating functional $W[\bar J, J] = \ln Z[\bar J, J]$, where $Z[\bar J, J]$ is obtained as a path integral of $\exp(-S[\bar \phi, \phi] + \bar J \bar \phi + J \phi)$ w.r.t.~the fields $\bar \phi$ and $\phi$. This allows to obtain an exact equation of motion for the density $\rho$ from the \emph{effective action} $\Gamma[\tpsi,\psi]$, the Legendre transform of  $W[\bar J, J]$, 
with the stationarity conditions
\acomment{
\begin{equation}
\label{eq:ofmotion}
	\delta \Gamma / \delta \psi = 0 = \delta \Gamma / \delta \tpsi  \quad \text{at } \tpsi = 0,\,\psi = \rho \,,
\end{equation}}
the macroscopic analog of the classical field equations. 
\rechange{Our analysis below will show that for the coagulation process the effective action $\Gamma$ takes a similar form as the microscopic action $S$. In fact, all but the term characterizing the coagulation process, $S_\lambda$, remain unaffected by fluctuations. For the latter, fluctuations renormalize the microscopic reaction kernel $\lambda(\mathbf{x})$ to its macroscopic counterpart $\mu(\mathbf{x})$. Moreover, fluctuations give rise to further contributions in the expansion of the effective action $\Gamma$, the most relevant of which is found to be proportional to $\int  \! \mathrm{d}t \sum_{\mathbf{x}} \tpsi(\mathbf{x},t) \psi^{\frac{5}{2}}(\mathbf{x},t)$.
Thus, to the order treated in this article, the `extremal principle', Eq.~(\ref{eq:ofmotion}), yields the kinetic equation  
\begin{equation}
\partial_t \rho =  - F (\rho) \, , 
\end{equation}
for states homogeneous in space.
The leading term of the \emph{non-equilibrium force}  $F(\rho)$ is the LMA term $\mu \rho^2$, where $\mu:=\sum_{\mathbf{x}}\mu(\mathbf{x})$ is the macroscopic decay rate. In addition, there are higher order terms in the density which violate the LMA, in particular a contribution $\sim \rho^{\frac{5}{2}}$, which is derived below.} 

The effective action can be calculated upon employing a non-perturbative RG analysis based on the Wetterich flow Eq.~\cite{Berges:2002p16426,Canet:2004p18293,Canet:2004p83,Canet:2006p16432}
\acomment{\begin{equation*} \label{eq:wetter}
	\partial_k \Gamma_k [\tpsi,\psi] = \frac{1}{2} \text{Tr}\! \left[\partial_k \hat R_k \left(\hat\Gamma^{(2)}_k[\tpsi,\psi] + \hat R_k \right)^{-1}\right]    \,,
\end{equation*}}
with the flow parameter $k$. The equation connects the microscopic action $S[\tpsi,\psi] \equiv \Gamma_{k=\infty}[\tpsi,\psi]$ with the macroscopic, effective action  $\Gamma[\tpsi,\psi]\equiv \Gamma_{k=0}[\tpsi,\psi]$, where all modes are integrated. This is mediated by the cutoff term $\hat R_k$ suppressing modes with momentum $q^2<k^2$ while not affecting those with $q^2>k^2$. The trace Tr runs over the function space of $\tpsi$ and $\psi$. $\hat \Gamma_k^{(2)}$ and $\hat R_k$ denote the $2 \times 2$ matrices of the second functional derivatives of $\Gamma_k$ and of the mass term $\Delta S_k = \int_{\mathbf{q},\omega} R_k(q^2) \tpsi(-\mathbf{q},-\omega) \psi(\mathbf{q},\omega)$, respectively. 

\acomment{\begin{figure}
\centering
    \mbox{\includegraphics[width=0.487\textwidth]{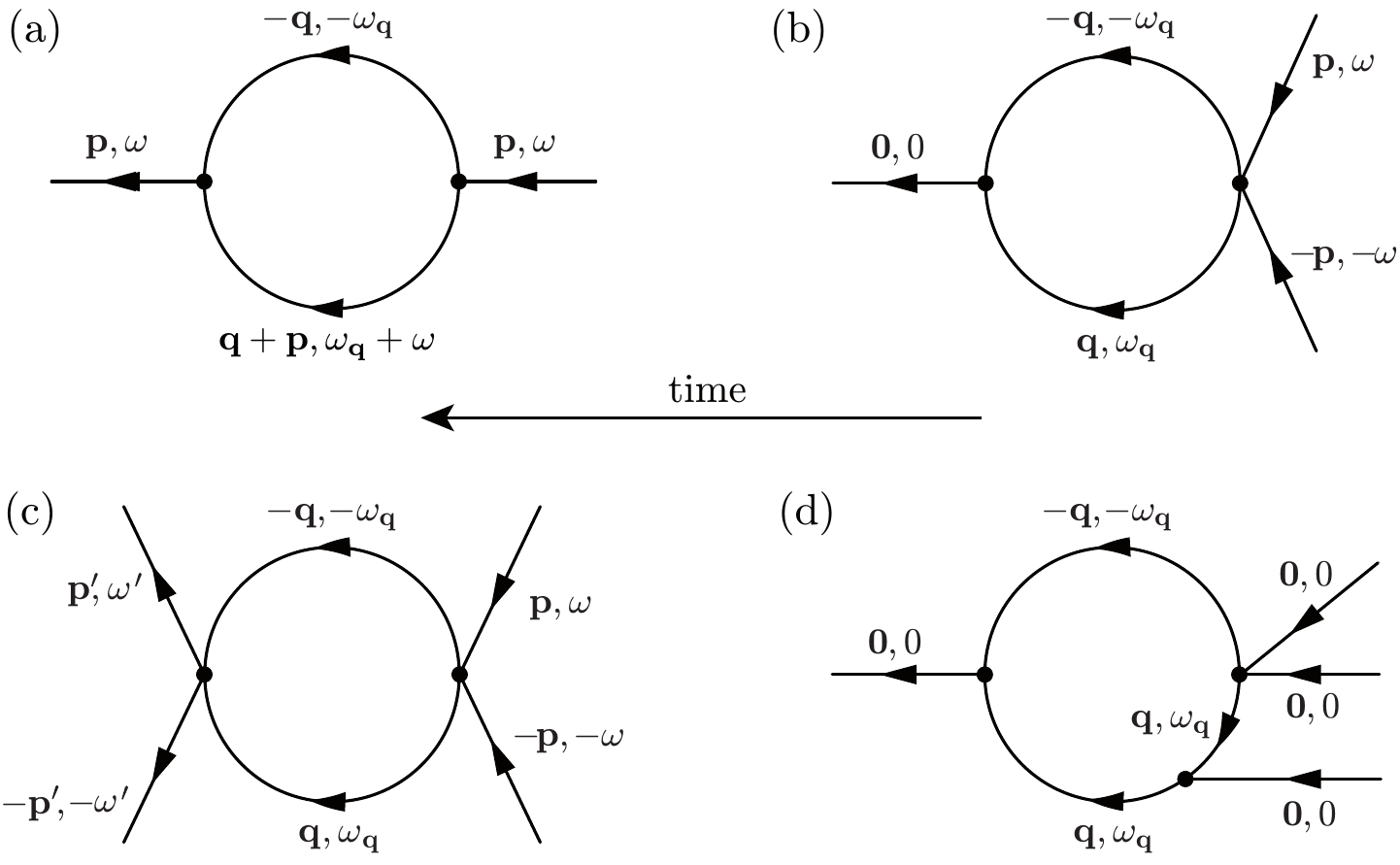}}
    \caption{Important one-loop Feynman diagrams.  
 Due to causality the propagator only connects earlier $\tpsi$ (`creation') to later $\psi$ (`annihilation'). Since for coagulation processes the number of legs can only decrease as time passes, the set of possible diagrams is substantially restricted. Indeed, there is no such diagram as in (a),  and, therefore, no renormalization of the dispersion $\epsilon$ and of the field amplitude $Z$. The diagrams depicted in (b) and (c) both stand for the renormalization of the reaction kernel $\lambda_k$. 
Finally, diagram~(d) exemplifies the divergence of the RG flow contribution of one-loop diagrams which only contain (2,1)- and (2,2)-vertices: Each of the three propagators gives rise to a factor $\sim 1/k^2$. For small $k$, the volume of reciprocal space and time that is integrated is $\sim k^{5}$. 
Thus, $g_k^{(1,3)} \sim k^{-1}$.}
\label{fig:oneloop}
\end{figure}\acomment}

We first apply the RG procedure to calculate the classical LMA term, before discussing deviations from that law. To this end, we take as an ansatz a `minimal' truncation for the effective average action,
\acomment{\begin{equation}
\label{eq:truncation}
	\Gamma_k[\tpsi,\psi] = S_{\lambda_k}[\tpsi,\psi]  + S_{\epsilon_k}[\tpsi,\psi] + S_{Z_k} [\tpsi,\psi]  \,,
\end{equation}}
which only contains terms already present in the initial action. Here $S_{Z_k} [\tpsi,\psi] = Z_k \int \! \mathrm{d}t \sum_{\mathbf{x}} \tpsi \partial_t \psi$, 
$\lambda_k$ is the renormalized reaction kernel and $\epsilon_k$ the renormalized dispersion relation. 
Though the RG flow generates higher order terms outside the functional space defined by the minimally truncated effective action, the flow of $\epsilon_k$, $Z_k$, and, under certain condition, also of $\lambda_k$, remains unaffected.
This can be shown upon recasting the Wetterich equation in a form amenable to a diagrammatic analysis: $\partial_k \Gamma_k  =   \tilde{\partial}_k \mathcal{D}_k$, where $\mathcal{D}_k = \frac{1}{2} \text{Tr} \ln ( \hat\Gamma_k^{(2)}  + \hat R_k )$ and $\tilde{\partial}_k$ acts on the $k$-dependence of $\hat R_k$ only. Familiar from perturbation theory, $\mathcal{D}_k$ creates the one-loop Feynman diagrams to the $(m,n)$-vertex functions $\Gamma_k^{(m,n)}$ with propagator $1/ ( \hat\Gamma_k^{(2)}  + \hat R_k )$. 
\change{This flow-equation approach to RG must be integrated with the full vertex functions and propagators, which reconstructs all loop corrections.}

For coagulation processes, the number of legs in the Feynman diagrams can only decrease as time passes. 
Hence certain processes, e.g.~corresponding to the diagram shown in Fig.~1(a), are physically not allowed, leading to a drastic restriction in possible diagrams. 
As a consequence, similar to the absence of propagator renormalization in perturbative RG~\cite{Peliti:1986p11725}, the dispersion relation and the field amplitude are not renormalized, $\epsilon_k (\mathbf{q}) = \epsilon (\mathbf{q})$ and $Z_k = 1$. 
The diagrams depicted in Fig.~1(b,c) determine the renormalization of 
the $\lambda_k\tpsi \psi^2$ and $\lambda_k\tpsi^2 \psi^2$ term, respectively, characterizing the coagulation process. 
As external legs do not contribute and the internal momenta and frequencies are independent of the external ones, the diagrams 
give equal contributions and $\lambda_k$ is well-defined. 
Moreover, since the diagram (Fig.~1(c)) for the renormalization of the (2,2)-vertex function involves only (2,2)-vertices, one obtains a closed, analytic solution for $\lambda_k$.
Taking the cutoff mass $R_k(\mathbf{q}) = (k^2 - \epsilon(\mathbf{q})) \, \Theta (k^2 - \epsilon(\mathbf{q}))$~\cite{Dupuis:2008p12392} yields the flow equation for the reaction kernel
\acomment{
\begin{equation}
	\label{eq:lambda_flow_x}
	\partial_k \lambda_k(\mathbf{x}) = \frac{2 \lambda_k(\mathbf{x}) \left(\mathcal{P}\circ\lambda_k\right)\!(\mathbf{x})}{k^3} \,,
\end{equation}}
with the projection $\left(\mathcal{P} \circ\lambda_k\right)\!(\mathbf{x}) = \int_\mathbf{q} \exp( i \mathbf{q} \cdot \mathbf{x})  \lambda_k(\mathbf{q}) \cdot \Theta(k^2 - \epsilon(\mathbf{q}))$. 
For many reaction kernels this 
simple equation is exact 
and allows to analyze general coagulation processes by calculating the macroscopic decay rate $\mu=\sum_{\mathbf{x}}\lambda_{k=0}(\mathbf{x})$. 
\change{In particular, this applies to spheres in continuum space and particles covering only a single site on a lattice (``one-site objects''). \rechange{This and the derivation of Eq.~(\ref{eq:lambda_flow_x}) is detailed in the Supplementary Material.}}
Actually, it is possible to derive a slightly more complex formula, which is exact for all reaction kernels~\cite{Winkler:2011}.

\acomment{\begin{figure}
\centering
    \mbox{\includegraphics[width=0.48\textwidth]{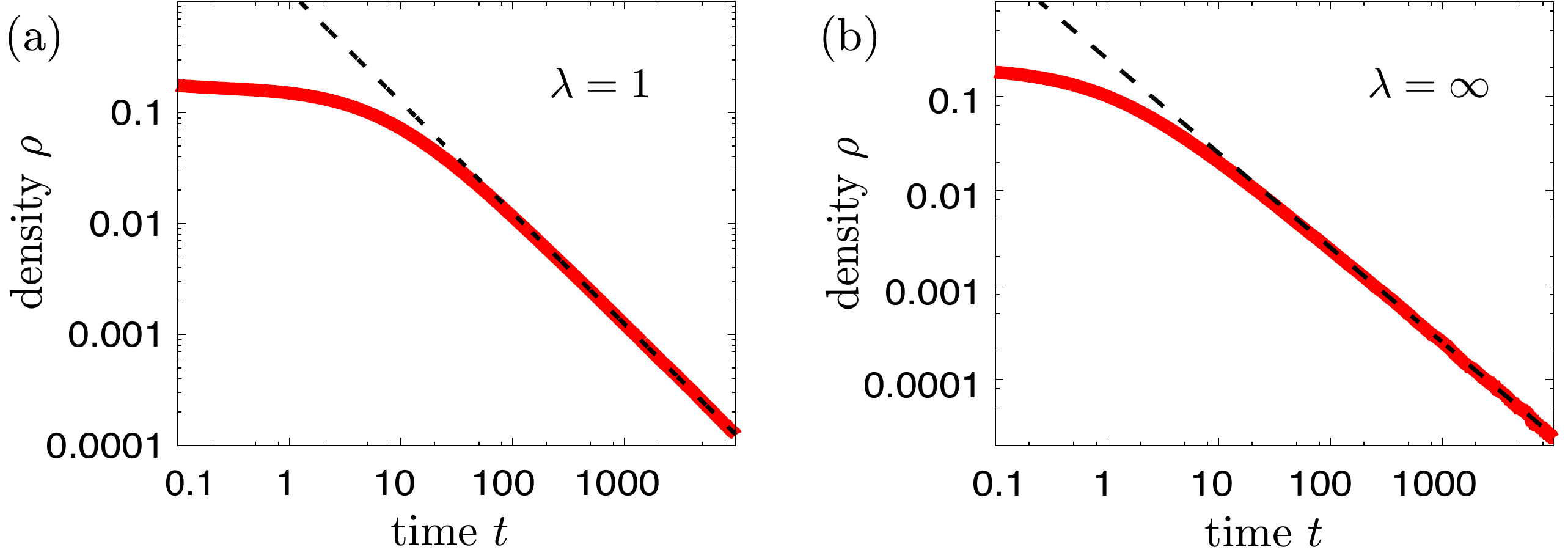}}
    \caption{(color online) Relaxation of the density for one-site objects.  On double logarithmic plots, data of stochastic simulations (solid red) are compared to the theoretical prediction for the density decay $\rho(t) \sim \mu^{-1}  t^{-1}$ (dashed line), with the macroscopic decay rate $\mu^{-1} = \lambda^{-1} +0.253\ldots$, cf.~Eq.~(\ref{eq:micro-macro_lattice}). The initial states in both plots were randomly distributed with $\rho(0) = 0.2$.}
\label{fig:relax}
\end{figure}}
\change{For one-site objects, with local interactions,} 
the flow equation reduces to
$\partial_k \lambda_k = 2 \lambda_k^2 \int_\mathbf{q} \Theta(k^2 - \epsilon(\mathbf{q}))  / k^3$,  with $\lambda_k:=\lambda_k(\mathbf{x}=\mathbf{0})$. We thus obtain an exact relation 
\acomment{\begin{equation}
	\frac{1}{\mu}  = \frac{1}{\lambda_0} = \frac{1}{\lambda} +  \int_{\mathbf{q}} \frac{1}{\epsilon(\mathbf{q})}  \,,
	\label{eq:micro-macro_lattice}
\end{equation}}
that connects the microscopic decay rate $\lambda$ with its macroscopic analog $\mu$ by a term that depends on the lattice structure via the dispersion relation $\epsilon$. 
Numerical integration yields
$\mu^{-1} = \lambda^{-1} + 0.252731009858(3)$, 
in excellent agreement with our stochastic simulations, cf.~Fig.~2. Eq.~(\ref{eq:micro-macro_lattice}) 
is not only valid for the cubic lattice, but for 
all Bravais lattices by inserting the corresponding dispersion relation. By the same token one can treat anisotropic diffusion and arbitrary dimension \change{$d>d_c$.} 

\acomment{\begin{figure}
\centering
    \mbox{\includegraphics[width=0.41\textwidth]{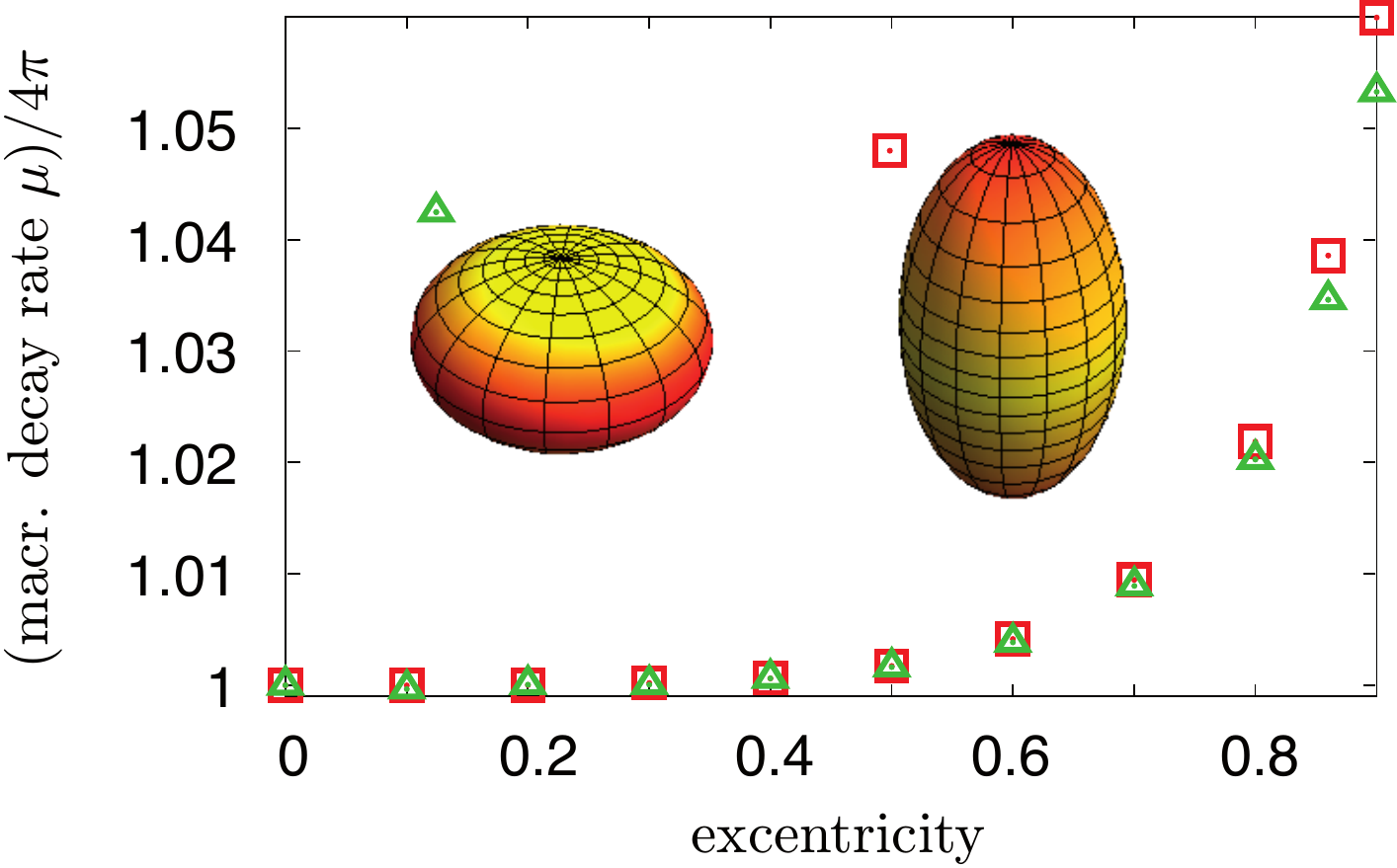}}
    \caption{ (color online) 
    Numerical solution to the flow equation, Eq.~(\ref{eq:lambda_flow_x}), for spheroids of volume $\frac{4}{3}\pi$ as a function of their eccentricity. 
   We observe that the macroscopic decay rate $\mu$ 
   increases with the eccentricity  and is larger for prolates. In the images, as yellow (light gray) turns to orange and red (dark gray), the $\mathbf{x}$-dependent macroscopic rate $\lambda_0(\mathbf{x})$ grows. As a rule of thumb, the more jagged the reaction kernel, the higher $\mu$ becomes.
}
\label{fig:spheroids}
\end{figure}}
The flow equation, Eq.~(\ref{eq:lambda_flow_x}), for the reaction kernel can also be applied to study the reaction kinetics of objects in continuous space, by simply considering the limit where the lattice spacing goes to zero.
This allows us to verify Smoluchowski's result, who studied spheres that coagulate instantaneously. 
Indeed, as detailed in the Supplementary Material, we recover his result $\mu = 4 \pi R$~\cite{Smoluchowski:1917p16756}, with $R$ the radius of the reaction kernel.  
To further illustrate the potential and versatility of our approach, 
we have solved Eq.~(\ref{eq:lambda_flow_x}) numerically for spheroids of equal volume $\frac{4}{3}\pi$. 
We find that the largest values of the macroscopic rates $\lambda_0(\mathbf{x})$ are attained at the sharp ends and edges of the prolates and oblates, respectively. This can be traced back to the fact that large momenta $q^2 > k^2$ do not contribute to the projection $\mathcal{P}$.

Finally, we extend our results beyond the lowest order in the density and discuss deviations from the LMA. To this end we represent the non-equilibrium force as a power series 
\begin{equation*}
	\rechange{F(\rho) = \lim_{k\to 0}  \sum_{n\ge 2} g_k^{(1,n)} \rho^n \,,}
\end{equation*}
exploiting the fact that the effective average action $\Gamma_k$ is analytic if $k>0$~\cite{Berges:2002p16426}.
For small densities we recover $F(\rho) = \mu \rho^2$ with the macroscopic decay rate $\mu = g_0^{(1,2)}$. 
The flow of the coefficients $g_k^{(1,n)}$ is determined by diagrams with $n$ incoming and one outgoing leg. Surprisingly, the most relevant term beyond $g_0^{(1,2)} \rho^2 = \mu \rho^2$ is not $g_0^{(1,3)} \rho^3$ as one might naively expect. In fact, in three dimensions \emph{all} coefficients $g_k^{(1,n)}$  ($n > 2$) turn out to diverge as $g_k^{(1,n)} \sim k^{5-2 n}$ for  $k \to 0$. (This follows from power counting, as illustrated in Fig.~1(d).)
Therefore, the infinite sum of diverging terms must scale as  
\begin{equation*}
	\rechange{\sum_{n\ge 3} g_k^{(1,n)}\rho^n \sim k^{5} f\left(\frac{\rho}{k^2}\right)\,,}
\end{equation*}
for some scaling function $f$. Since for large systems, the non-equilibrium force must become independent of the system size, i.e. independent of $1/k$, one obtains 
$f(x) \sim x^{\frac{5}{2}}$. This adds a non-analytic term $\sim\rho^{\frac{5}{2}}$ to the non-equilibrium force $F$.

\acomment{\begin{figure}
\centering
    \mbox{\includegraphics[width=0.41\textwidth]{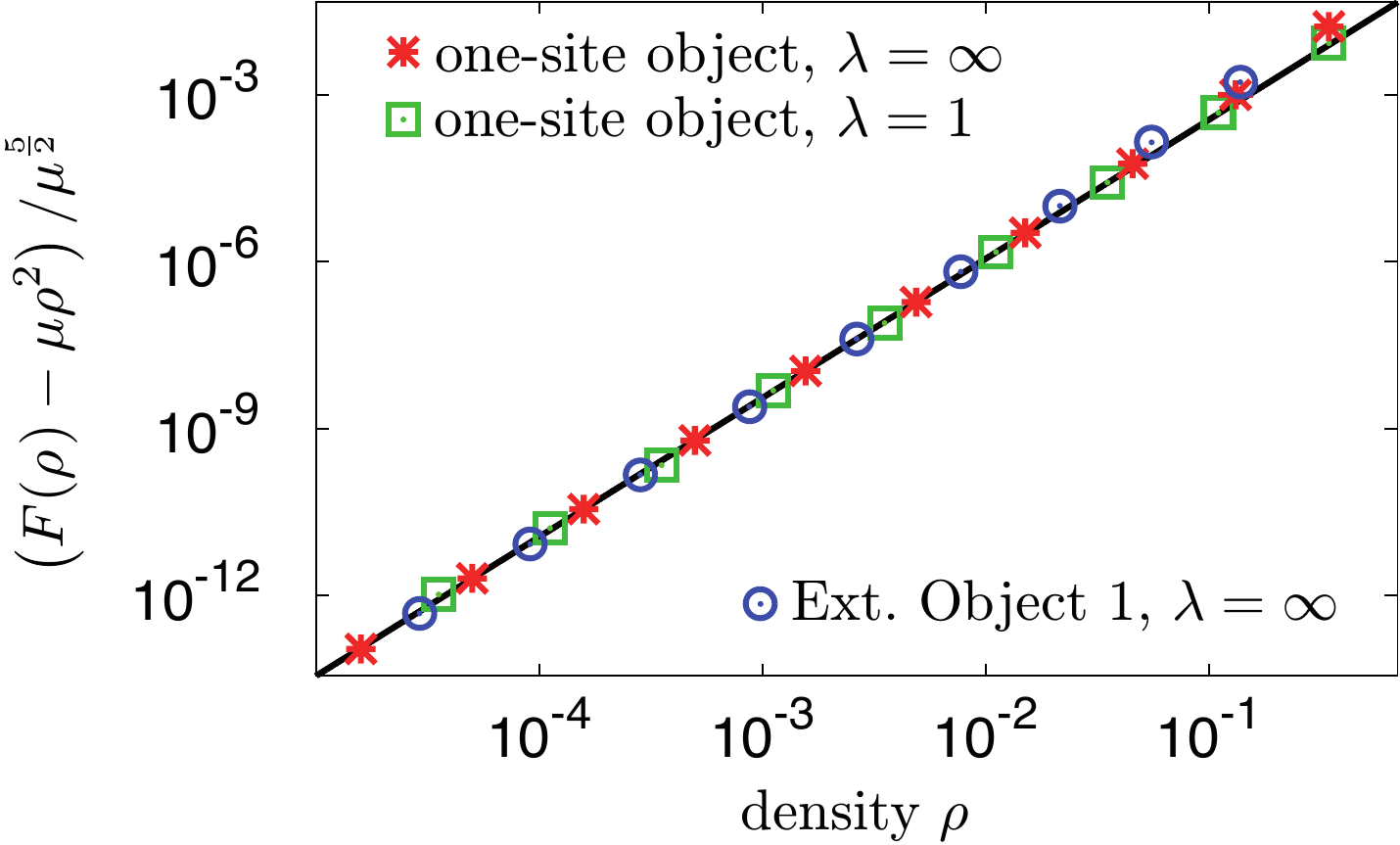}}
    \caption{(color online) Rescaled data for the deviation from the LMA. We ran simulations for a range of models and compared the results with the predicted universal correction $\rho^{\frac{5}{2}}/\left(2 \sqrt{2} \pi\right)$ (solid black line), cf.~Eq.~(\ref{eq:correction}).  
\rechange{The non-equilibrium force $F$ can be determined directly by introducing homogeneous particle input with rate $J$. This gives rise to an additional term $J$ in the kinetic equation, such that for the stationary state $F(\rho) = J$.} 
Extended Object 1 is a discretization of the sphere, made up of seven sites. Together with Extended Object 2, it is discussed in the Supplementary Material, where also the data for Extended Object 2 are provided (not shown here for clarity but in excellent agreement with the theory as well). 
 }
\label{fig:correction}
\end{figure}}

The divergent terms in the non-equilibrium force, originating in long-wavelength fluctuations, cannot resolve the reaction kernels and the lattice structure. As elaborated in the Supplementary Material, this can be exploited to calculate the $\rho^{\frac{5}{2}}$ term \emph{exactly} from the Wetterich equation. Overall we find for the non-equilibrium force (exact up to higher orders in $\rho$):
\acomment{\begin{equation}
\label{eq:correction}
	F(\rho) = \mu \rho^2 + \frac{\mu^{\frac{5}{2}}}{2 \sqrt{2} \pi}\rho^{\frac{5}{2}}  \,.
\end{equation}}
This equation bears a new fundamental insight: Beyond the LMA term, quadratic in the density, the non-equilibrium force driving the reaction kinetics contains a non-analytic term violating the LMA. Similar as for critical phenomena, long-wavelength fluctuations are the physical origin of this term. Unlike in critical dynamics, the anomalous power-law is not governed by an RG flow close to a fixed point but is a genuine strong coupling result. In contrast to low-dimensional systems the three-dimensional coagulation process is not critical. Nevertheless, we find that the term violating the LMA is a \emph{universal} function of the macroscopic decay rate $\mu$. From our theoretical analysis we anticipate this to be a generic feature of reaction processes in three dimensions with upper critical dimension $d_c = 2$.
We have run simulations for a  range of 
models (one-site objects with both finite and infinitely large reaction rates, and two examples of extended objects that react immediately on contact), cf.~Fig.~4, which clearly corroborate our theoretical findings.

Exciton luminescence has been previously used to investigate low-dimensional reaction kinetics: By accurate measurements on the fusion of excitons, anomalous  behavior was observed in an effectively one-dimensional system~\cite{BLAKLEY:1990p17131,Kroon:1993p15523}.  We expect that our prediction of a strong violation of the LMA could be revealed with similar kinds of experiments for three-dimensional systems. In addition, we believe that our results will stimulate further theoretical and experimental activities to explore the fundamental implications of fluctuations on reaction kinetics, and to map out the range of validity of the LMA.

Financial support of Deutsche Forschungsgemeinschaft through the German Excellence Initiative via the program `Nanosystems Initiative Munich' (NIM) and through the SFB TR12 `Symmetries and Universalities in Mesoscopic Systems' is gratefully acknowledged. 


\onecolumngrid
\newpage

\vspace{0.5cm}

\subsection{\large{Supplementary Material}}

\vspace{0.5cm}

\subsubsection{The field theoretic action for coagulation.} 
In order to illustrate the Doi-Peliti formalism to readers unfamiliar with the subject matter, let us consider coagulation $A+A\to A$ with rate $\lambda$ in `zero' dimensions, i.e.~when there is no spatial structure. Its Master equation reads $\partial_t \, p_n(t) = \lambda (n+1) n \, p_{n+1}(t) - \lambda n (n-1) \, p_n(t)$, where $p_i$ denotes the probability of there being $i$ particles in the system. We observe that it can be recast as $\partial_t P(t) =  - \mathcal{H}(a^{+}, a^{-}) P(t)$, where $P(t) = \sum_n p_n (t) | n \rangle$ is the probability vector,  $\mathcal{H}(a^{+}, a^{-}) = \lambda \left(a^{+} a^{+}  a^{-} a^{-}  - a^{+} a^{-} a^{-} \right)$ is the stochastic Hamiltonian, and $a^{+}$ and $a^{-}$ are the ladder operators for creation and annihilation of particles, respectively. They obey $a^{+} | n \rangle = | n+1 \rangle$, $a^{-} | n\rangle = n | n - 1 \rangle$. 
In analogy to quantum mechanics one can pass to a field theory action $S$. The reaction part $S_\lambda$ of the action is obtained by replacing $a^{+}$ with $\hat \psi(t)$ and $a^{-}$ with $\psi(t)$ in the stochastic Hamiltonian $\mathcal{H}$ and integrating over time. 
\change{
Due to probability conservation $\hat \psi = 1$ for a reaction-diffusion system~\cite{Canet:2004p18293}. It is adequate to perform a field shift $\hat \psi \to 1 + \tpsi$. This is the underlying reason why the Gibbs functional $\Gamma[\tpsi,\psi]$ displays a saddle point at $\tpsi = 0$.
After adding a term for the time evolution, one finds the full action $S[\tpsi,\psi]  = \lambda \int \! \mathrm{d}t  \left(\tpsi(t)+1\right)\tpsi(t) {\psi(t)}^2 + \int \! \mathrm{d}t \, \tpsi(t) \partial_t \psi(t)$.
}

\subsubsection{Exact flow equation for the decay rate of one-site objects.}   
\begin{figure}
      \centering \includegraphics[width=0.8\textwidth]{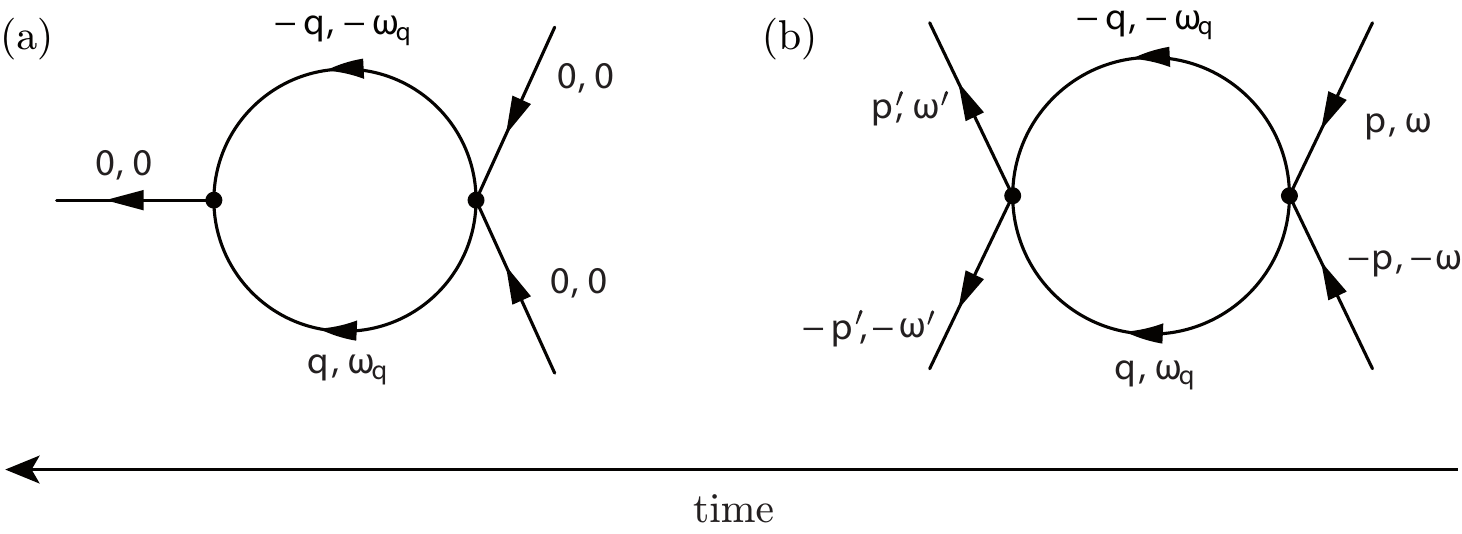}
\caption{One-loop Feynman diagrams for the calculation of the flow of the decay rate $\lambda_k$. Diagram (b) for the renormalization of the (2,2)-vertices only contains (2,2)-vertices. Thus one obtains a closed and exact formula.}
\label{fig:feyn}
\end{figure}

The microscopic action reads, 
\begin{IEEEeqnarray*}{rCl}
S[\tpsi,\psi]  &=& \int \! \mathrm{d}t \sum_{\mathbf{x}} \left( \lambda \left(\tpsi(\mathbf{x},t) + 1\right) \tpsi(\mathbf{x},t) \psi(\mathbf{x},t)^2 +  \tpsi(\mathbf{x},t) \partial_t \psi(\mathbf{x},t) \right) + \nonumber 
\\
  & & +\:   \int_{\mathbf{q},\omega}   \epsilon(\mathbf{q})\tpsi(-\mathbf{q},-\omega) \psi(\mathbf{q},\omega)\,. 
\end{IEEEeqnarray*}
$(m,n)$-vertex functions $\Gamma^{(m,n)}_{k\,(\mathbf{p}_1,\omega_1;\ldots;\mathbf{p}_{m+n},\omega_{m+n})}$ (related to one-loop Feynman diagrams with $n$ incoming and $m$ outgoing legs) are obtained by taking the functional derivative w.r.t.~$\tpsi(\mathbf{p}_i,\omega_i)$, $i \in \{1,\ldots, n\}$ and $\psi(\mathbf{p}_{j},\omega_{j})$, $j\in \{m+1,\ldots, m+n \}$. 
The renormalized reaction rate $\lambda_k$ is given by
\begin{equation*}
\label{eq:lambda}
	\lambda_k = \frac{(2\pi)^{12} \,\Gamma^{(1,2)}_{k\,(\mathbf{0},0;\mathbf{0},0;\mathbf{0},0)}}{V T} \,.
\end{equation*}
$V = \sum_{\mathbf{x}} = (2\pi)^3 \delta(\mathbf{0})$ and $T = \int \! \mathrm{d}t = 2\pi \delta(0)$ denote the asymptotically large volumes in space and time which are summed and integrated over, respectively (in the final result they drop out).   
The flow of 
$\Gamma^{(1,2)}_{k\,(\mathbf{0},0;\mathbf{0},0;\mathbf{0},0)}$ 
is associated with the graph in Fig.~\ref{fig:feyn}a, which includes, in addition, a (2,2)-vertex function with non-zero momenta and frequencies (right vertex of the graph). Let us therefore also consider the flow of 
$\Gamma^{(2,2)}_{k\,(\mathbf{p^\prime},\omega^\prime;-\mathbf{p^\prime},-\omega^\prime;\mathbf{p},\omega;-\mathbf{p},-\omega)}$, 
cf.~Fig.~\ref{fig:feyn}b. Evidently, it is a closed function of itself. The vertices and the internal momenta and frequencies are independent of the external ones. Since this also holds for the initial conditions, we have 
$\Gamma^{(2,2)}_{k\,(\mathbf{p}^\prime,\omega^\prime;-\mathbf{p}^\prime,-\omega^\prime;\mathbf{p},\omega;-\mathbf{p},-\omega)}  = \Gamma^{(2,2)}_{k\,(\mathbf{0},0;\mathbf{0},0;\mathbf{0},0;\mathbf{0},0)}$, for all $k$. 
Similarly, the identity $\Gamma^{(2,2)}_{k\,(\mathbf{0},0;\mathbf{0},0;\mathbf{0},0;\mathbf{0},0)} = \frac{1}{(2 \pi)^{4}} \Gamma^{(1,2)}_{k\,(\mathbf{0},0;\mathbf{0},0;\mathbf{0},0)}$, which holds initially, is inherited to all scales $k$. Thus, 
\[
	\Gamma^{(2,2)}_{k\,(\mathbf{p}^\prime,\omega^\prime;-\mathbf{p}^\prime,-\omega^\prime;\mathbf{p},\omega;-\mathbf{p},-\omega)} = \frac{V T}{(2\pi)^{16}} \lambda_k\,.
\]
The evaluation of the flow of $\lambda_k$ through the  diagram in Fig.~\ref{fig:feyn}b at zero momenta and frequencies therefore involves no approximation.  With cutoff mass $R_k(\mathbf{q}) = (k^2 - \epsilon(\mathbf{q})) \, \Theta (k^2 - \epsilon(\mathbf{q}))$ it yields (recall that $\tilde \partial_k = \partial_k R_k \cdot \partial_{R_k}$)
\begin{equation*}
	\partial_k \lambda_k = 
	- 4 \frac{1}{2} \tilde\partial_k \int_{\mathbf{q},\omega} \lambda_k \frac{1}{R_k(\mathbf{q}) + \epsilon(\mathbf{q}) - i \omega} \lambda_k \frac{1}{R_k(\mathbf{q}) + \epsilon(\mathbf{q}) + i \omega} = 
	2 \lambda_k^2 \frac{\int_\mathbf{q} \Theta(k^2 - \epsilon(\mathbf{q}))}{k^3}\,.
\end{equation*}
Thus the exact solution for the macroscopic decay rate is 
\begin{equation}
\label{eq:onesite}
	\frac{1}{\mu}  = \frac{1}{\lambda_0} = \frac{1}{\lambda} +  \int_{\mathbf{q}} \frac{1}{\epsilon(\mathbf{q})} \,.
\end{equation}

\change{We remark that for this particularly simple example one can arrive at the same result  by summing over a sequence of bubble diagrams, which also plays an important role in the calculation for spheres in \cite{Doi:1976p15932,Mikhailov:1985p19587}. One needs to consider 
\begin{figure}[H]
      \centering \includegraphics[width=0.99\textwidth]{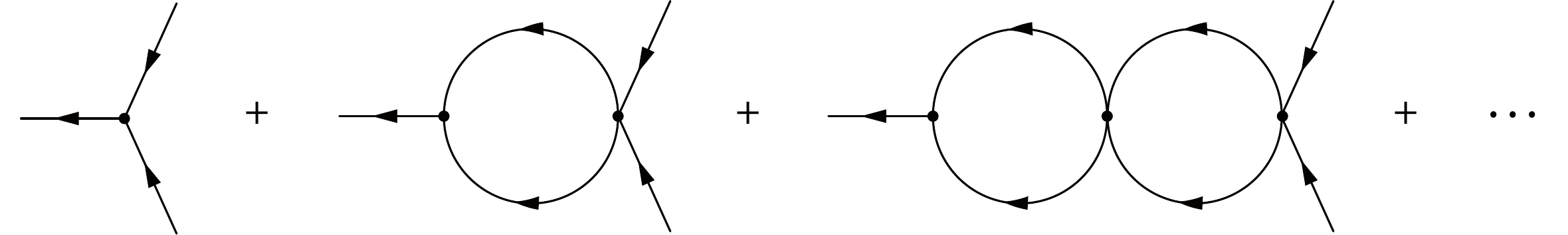}
\end{figure}
\noindent For one-site objects these bubble diagrams factorize, such that
\[
	\mu = 2 \lambda +  (-\lambda) \int_{\mathbf{q},\omega} \frac{1}{\epsilon(q)^2 + \omega^2}  2 \lambda +  \left[(-\lambda) \int_{\mathbf{q},\omega} \frac{1}{\epsilon(q)^2 + \omega^2}\right]^2 2 \lambda+ \ldots 
\]
which confirms Eq.~(\ref{eq:onesite}).}

\subsubsection{Flow equation for the decay rate of extended objects.}  
With the truncation of the main text, in Fourier space the reaction part of the effective average action $\Gamma_{k}[\tpsi,\psi]$ reads 
\begin{IEEEeqnarray*}{C}
	 \int_{\mathbf{p},\omega,\mathbf{p^\prime},\omega^\prime,\mathbf{q},\omega^{\prime\prime}} \lambda_k(\mathbf{q}) \left( \tpsi(-\mathbf{p}-\mathbf{q},-\omega-\omega^\prime - \omega^{\prime\prime}) + (2\pi)^4 \delta(-\mathbf{p}-\mathbf{q})\delta(-\omega-\omega^\prime - \omega^{\prime\prime}) \right) \cdot	 \\
		\cdot  \tpsi(-\mathbf{p}^\prime + \mathbf{q},\omega^{\prime\prime}) \psi(\mathbf{p}^\prime,\omega^\prime) \psi(\mathbf{p},\omega) \,.
\end{IEEEeqnarray*}
The renormalized reaction kernel is given by
\begin{equation*}
	\lambda_k(\mathbf{p}) =  \frac{(2\pi)^{12} \,\Gamma^{(1,2)}_{k\,(\mathbf{0},0;\mathbf{p},0;-\mathbf{p},0)}}{V T}\,.
\end{equation*}
In analogy to one-site objects, 
\begin{equation*}
	\Gamma^{(2,2)}_{\mathbf{k\,(p^\prime},\omega^\prime;\mathbf{p^\prime},\omega^\prime;-\mathbf{p^\prime},-\omega^\prime;\mathbf{p},\omega;-\mathbf{p},-\omega)} 
	= \frac{V T}{(2 \pi)^{16}} \lambda(\mathbf{p}+\mathbf{p}^\prime) \,.
\end{equation*}
Thus evaluating the flow of $\lambda_k(\mathbf{p})$, cf.~Fig.~\ref{fig:feyn}b for $\mathbf{p}^\prime = \mathbf{0}$,  yields
\begin{equation}
	\partial_k \lambda_k(\mathbf{p}) = 2 \int_{\mathbf{q}} \, \lambda_k(\mathbf{p}-\mathbf{q}) \lambda_k(\mathbf{q}) \frac{\Theta(k^2 - \epsilon(\mathbf{q}))}{k^3}\,.
	\label{eq:lambda_flow_p}
\end{equation}
Equivalently, 
\begin{equation}
	\label{eq:lambda_flow_x}
	\partial_k \lambda_k(\mathbf{x}) = \frac{2 \lambda_k(\mathbf{x}) \left(\mathcal{P}\circ\lambda_k\right)\!(\mathbf{x})}{k^3}\,.
\end{equation}
with the projection $\left(\mathcal{P}\circ\lambda_k\right)\!(\mathbf{x}) = \int_{\mathbf{q}} \exp( i \mathbf{q} \cdot \mathbf{x})  \lambda_k(\mathbf{x}) \Theta(k^2 - \epsilon(\mathbf{q}))$.

\begin{figure}
      \centering \includegraphics[width=0.6\textwidth]{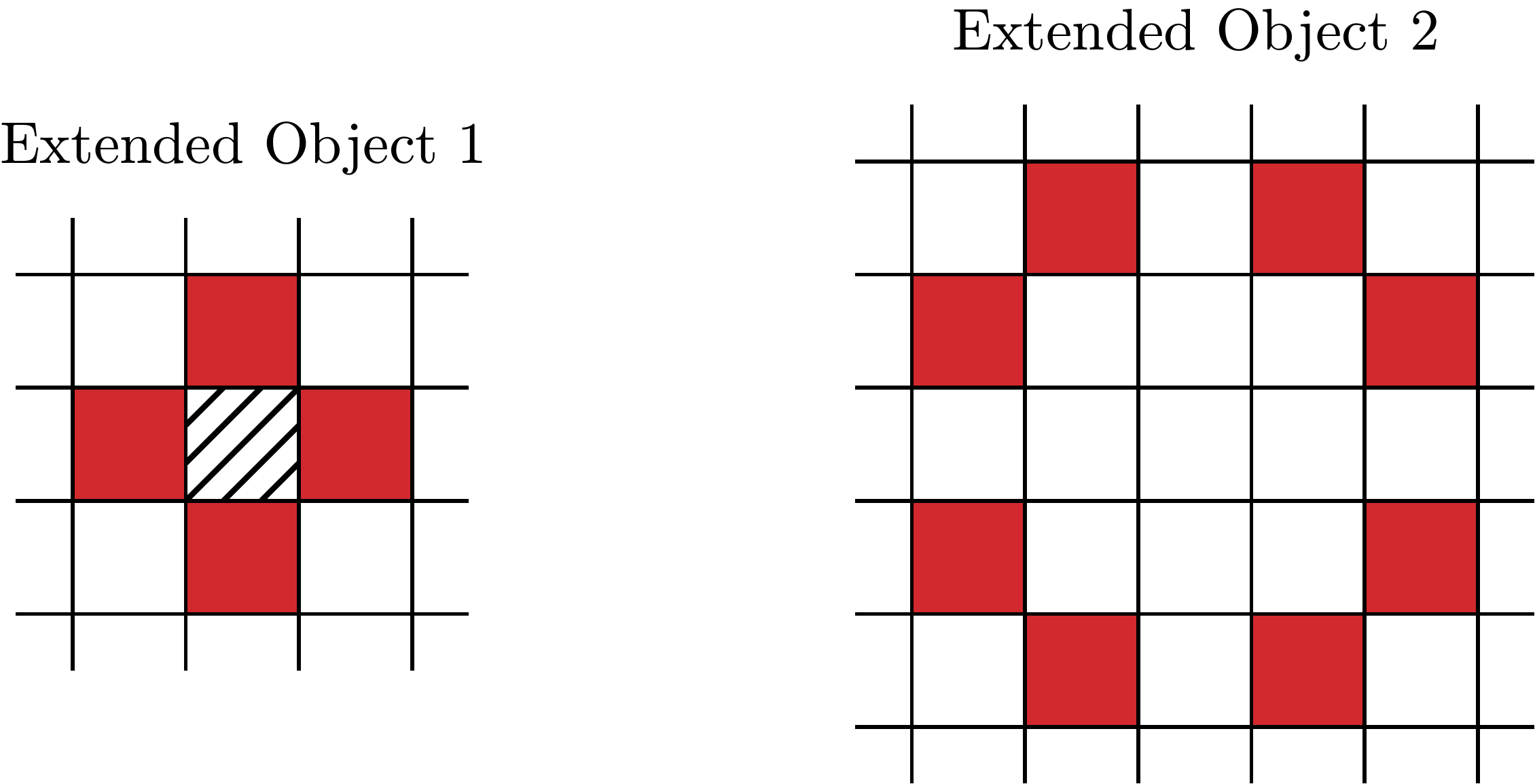}
\caption{Projection of extended objects in two dimensions (solid red). 
The  three-dimensional version of Extended Object 1 is obvious, for Extended Object 2 it is created by the sites $(1,1,2),(1,2,1),(2,1,1)$ and their mirror image in each octant. They encompass 6 and 24 sites, respectively. For instantaneous reactions, the striped square  can be regarded as part of Extended Object 1.}
\label{fig:extended_objects}
\end{figure}
\subsubsection{Numerical calculation of the macroscopic decay rate for two selected, extended reaction kernels.} 
As long as the support of the reaction kernel is finite, the flow Eq.~(\ref{eq:lambda_flow_x})  reduces to a finite set of coupled ordinary differential equations, with dimension equal to the number of sites that make the reaction kernel. Here we treat reaction kernels which allow for a particularly precise numerical solution, because they can be described by only one degree of freedom. Incidentally, this is the underlying reason why the flow equation for the LMA term is exact, as proven in a forthcoming publication. Let us consider the two reaction kernels whose two-dimensional versions are depicted in Fig.~\ref{fig:extended_objects}.  
The lattice sites that these objects consist of (i.e. the sites that are part of the reaction kernel) are all equivalent. This symmetry must hold along the renormalization group flow. Furthermore, the flow equation conserves the support of the reaction kernel in position space. Thus the objects also keep their shape. 
Explicitly, in three dimensions the renormalized reaction kernel of Extended Object 1 can be expressed as
\begin{equation*}
	\lambda_k(\mathbf{p}) = \tilde\lambda_k \sum_{\nu = 1}^3 \left( e^{+ i p_\nu}  + e^{-i p_\nu}\right)  =  2 \tilde\lambda_k \sum_{\nu = 1}^3 \cos(p_\nu) \,,
\end{equation*}
The microscopic decay rate is $\lambda := \lambda_{\infty} (\mathbf{0}) = 6 \tilde\lambda_\infty$ ($\tilde \lambda_\infty$ is the reaction rate for each site of the kernel). Solving  Eq.~(\ref{eq:lambda_flow_p}) at $\mathbf{p}=\mathbf{0}$, one obtains the macroscopic decay rate
 \begin{IEEEeqnarray*}{C}
 	\frac{1}{\mu}  =  \frac{1}{\lambda_0(\mathbf{0})} = \frac{1}{\lambda_\infty(\mathbf{0})} +  \frac{1}{18} \int_0^\infty \!\mathrm{d}k \int_{\mathbf{q}}  \left( 2 \sum_{\nu=1}^3 \cos(q_\nu) \right)^2 \frac{\Theta\!\left(k^2 - \epsilon(\mathbf{q}) \right)}{k^3}  =  
	\nonumber \vspace{0.5em} \\
	 =  \frac{1}{\lambda} + \underbrace{\int_{\mathbf{q}}  \frac{\left(\sum_{\nu = 1}^3 \cos(p_\nu)\right)^2}{36 \sum_{\nu=1}^3 \sin^2(q_\nu/2)}}_{0.086064343192(3)}\,.
\end{IEEEeqnarray*}
In the last step we inserted the dispersion relation $\epsilon(\mathbf{p}) = 4 \sum_{\nu=1}^3 \sin^2(p_\nu/2)$ for  the cubic lattice with lattice spacing 1.
The calculation for Extended Object 2 is analogous. If each of the 24 sites effects a reaction with rate $\tilde\lambda_\infty$ then
\begin{equation*}
	\frac{1}{\mu} = \frac{1}{\lambda} + 0.036287603611(2)\,,
\end{equation*}
with the microscopic decay rate $\lambda = 24 \tilde\lambda_\infty$. The simulations confirm these results for the macroscopic decay rates and support our prediction on the universal correction to the non-equilibrium force, cf.~Fig.~\ref{fig:correction}. 
\begin{figure}
\centering
    \mbox{\includegraphics[width=0.55\textwidth]{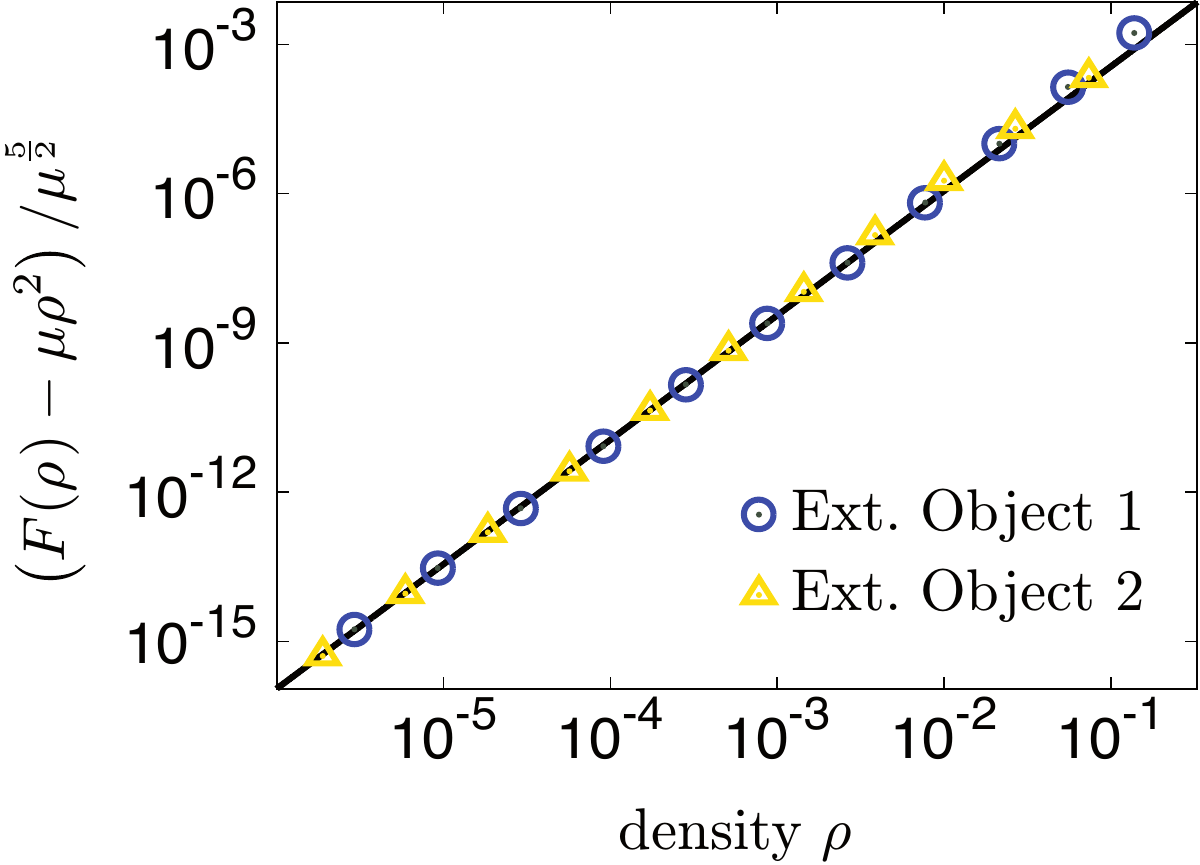}}
    \caption{Rescaled data for the universal correction to the non-equilibrium force for the extended objects. The data collapse well on the predicted universal correction $\frac{\rho^{\frac{5}{2}}}{2 \sqrt{2} \pi}$.  We considered instantaneous reactions $\tilde\lambda_\infty = \infty$.}
\label{fig:correction}
\end{figure}

\subsubsection{The macroscopic decay rate for instantaneously reacting spheres.}

We choose the reaction kernel to be the surface of a sphere,
\begin{equation*}
	\lambda_k(\mathbf{x}) = \tilde\lambda_k \, \delta(R - x)\,.
\end{equation*}
In the limit of an infinitely large microscopic reaction rate this is evidently identical with spheres that coagulate instantaneously on contact.
Without loss of generality, we set the radius $R = 1$ in the following. Using spherical coordinates, 
\begin{IEEEeqnarray*}{rCl}
	\left( \mathcal{P} \circ \lambda_k\right) (\mathbf{x}) & = & \frac{1}{(2 \pi)^3} \int \!\mathrm{d}q \ud\vartheta \ud\phi\, q^2\sin(\vartheta) e^{i q x \cos(\vartheta)} \Theta\!\left(k^2 - q^2\right) 
	\cdot  
	\nonumber \vspace{0.5em} \\
	 & &\cdot \:	\int \!\mathrm{d} r \ud \tilde\vartheta\ud\tilde\phi \, r^2 \sin(\tilde\vartheta) e^{-i q r \cos(\tilde\vartheta)} \tilde\lambda_k \,\delta(1-r) = 
	\nonumber \vspace{0.5em} \\
	 & = & \tilde\lambda_k \underbrace{\frac{2}{\pi} \int_0^k \!\mathrm{d} q\, \frac{1}{x} \sin(q x) \sin(q)}_{=: f_k(x)}\,.
\end{IEEEeqnarray*}
\begin{IEEEeqnarray*}{rCl}
	&  \Longrightarrow \quad&  \partial_k \tilde\lambda_k  = 2 \tilde\lambda_k^2 \frac{f_k(1)}{k^3} 
	 \nonumber \vspace{0.5em} \\
	& \Longrightarrow \quad & \frac{1}{\tilde\lambda_0}   =  \frac{1}{\tilde\lambda_\infty} + 2 \int_0^\infty \!\mathrm{d} k\, \frac{f_k(1)}{k^3} = \frac{1}{\tilde\lambda_\infty} +1\,.
\end{IEEEeqnarray*}
The macroscopic decay rate becomes
\begin{equation*}
	\frac{1}{\mu} = \frac{1}{\lambda} + \frac{1}{4\pi} \,, 
\end{equation*}
where $\lambda = 4\pi \tilde \lambda_\infty$. Therefore, 
\begin{equation*}
	\mu = 4 \pi D R\,,
\end{equation*}
with diffusion constant $D$ and radius $R$. This confirms Smoluchowski's result~\cite{Smoluchowski:1917p16756}, proved to be exact by Doi~\cite{Doi:1976p15932}.

\subsubsection{The universal term of the non-equilibrium force.}
To extract the amplitude of the $\rho^{\frac{5}{2}}$ term in the non-equilibrium force $F(\rho) = \lim_{k\to 0}  \sum_{n\ge 2} g_k^{(1,n)} \rho^n$ with $g_k^{(1,n)} \sim \tilde g_k^{(1,n)} k^{5-2n}$ , we first observe that $\tilde g^{(1,n)}$  are universal functions of the macroscopic decay rate $\mu=\lambda_{k=0}(\mathbf{q}=\mathbf{0})$: To leading order in $k$ we can set the argument $\mathbf{q} = \mathbf{0}$, because the cutoff cancels momenta larger than $k$, $\partial_k R_k(\mathbf{q}) = 0$ when $q^2 > k^2$, and also $k$ may be set to zero to lowest order in the calculation. In other words, the divergent terms only `see' structureless point particles embedded in continuous space. We choose the ansatz
\begin{equation}
	\Gamma_k[\tpsi,\psi] = \int \! \mathrm{d}t \sum_x U_k(\tpsi,\psi) + S_{\epsilon_k = \epsilon} + S_{Z_k = 1} \,,
\end{equation}
which includes all terms needed to calculate the correction exactly. The local potential $U_k$ is related to the non-equilbirium force via 
$F(\rho) = \partial_{\tpsi} U_0(\tpsi,\rho)|_{\tpsi=0}$.
Its flow equation is readily follows from the Wetterich equation. With our cutoff mass it reads
\begin{equation}
\label{eq:dUk}
	\partial_k U_k(\tpsi,\psi) =  \frac{\mathcal{V}(k)\,\, k\left( \frac{\partial^2 U_k(\tpsi,\psi)}{\partial\tpsi\partial\psi} + k^2\right)}{\sqrt{\left( \frac{\partial^2 U_k(\tpsi,\psi)}{\partial\tpsi\partial\psi}  + k^2\right)^2 - \frac{\partial^2 U_k(\tpsi,\psi)}{\partial\tpsi\partial\tpsi}  \frac{\partial^2 U_k(\tpsi,\psi)}{\partial\psi\partial\psi} }}\,, 
\end{equation}
where $\mathcal{V}(k) = \int_{\mathbf{q}} \Theta(k^2-\epsilon(\mathbf{q})) \sim \frac{4}{3} \pi \left(\frac{k}{2 \pi}\right)^3$. 

\change{
Focussing on the physically relevant terms, i.e.~terms proportional to $\tpsi$, cf.~the extremal principle, we can finally calculate the exact value of the universal correction to $F$. 
Since we are interested in long-range fluctuations, i.e.~small $k$, we may replace $\mathcal{V}(k) \to \frac{4 \pi}{3 (2 \pi)^3} k^3$. Substituting $U_k(\tpsi,\psi)\to \mu \tpsi^2 \tpsi^2 + \mu \tpsi \psi^2$ on the right side of  Eq.~(\ref{eq:dUk}) then gives for $F_k(\rho) = \frac{\partial U_k(\tpsi = 0,\psi=\rho)}{\partial \tpsi}$,
\begin{equation}
\label{eq:fluct}
	\partial_k  F_k(\rho) \approx \frac{\mu^2 k^4  \rho^2}{3 \pi^2 \left(k^2+2 \mu \rho \right)^2}\,.
\end{equation}	
Although this equation is not exact, according to the discussion in the main text it still yields the correct non-analytic contribution $\sim \rho^{\frac{5}{2}}$, because it treats all the terms which give rise to it exactly. 
Indeed, integrating Eq.~(\ref{eq:fluct}) from any $k = \Lambda > 0$ to $k=0$ yields a  contribution  $\mu^{\frac{5}{2}}/\left({2 \sqrt{2} \pi}\right) \rho^{5/2}$ to the non-equilibrium force $F$. Similarly, one can calculate terms of higher order in $F$. In particular we find that the next correction is of order $\rho^3 \ln(\rho)$, and is again given by a universal function, as detailed in a forthcoming publication. 

The non-analytic correction term also can be obtained by a perturbative calculation, as suggested by an anonymous referee. 
Assuming that the relevant diagrams which need to be resummed are of the form
\begin{figure}[H]
      \centering \includegraphics[width=0.27\textwidth]{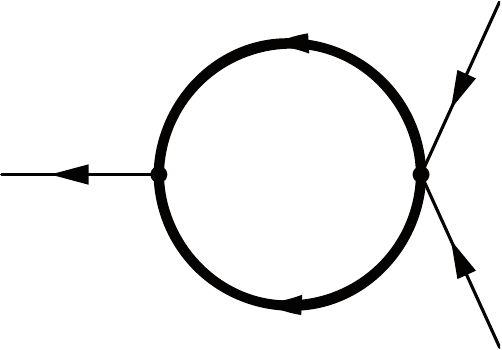}
\end{figure}
\noindent with the response function 
\begin{figure}[H]
      \centering \includegraphics[width=0.99\textwidth]{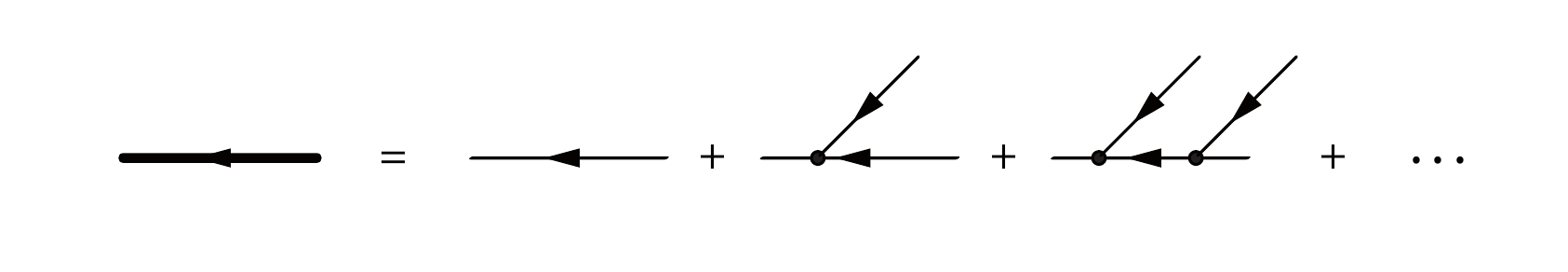}
\end{figure}
\noindent one finds  a contribution
\[
	- \int_{\mathbf{q}} \int_{t>0} \! \mathrm{d}t \, e^{- t (q^2 + 2 \mu \psi)} \tpsi \psi^2 \mu^2 \,.
\]
In writing this we have replaced the vertices by their macroscopic value $\mu$. Thus again there arises the correction $\mu^{\frac{5}{2}}/\left({2 \sqrt{2} \pi}\right)  \rho^{5/2}$ to the non-equilibrium force $F$.}

\end{document}